**Using Fluorescence Recovery After Photobleaching (FRAP) to study dynamics of the Structural Maintenance of Chromosome (SMC) complex *in vivo***

Anjana Badrinarayanan[1,3] and Mark C. Leake[2]

[1] Department of Biology

Massachusetts Institute of Technology

Cambridge

MA 02139

USA

[2] Biological Physical Sciences Institute (BPSI)

University of York

York

YO10 5DD

United Kingdom

[3] Corresponding author

e–mail: anjana1@mit.edu

Tel: 617-253-3677

Running head: FRAP as a tool to study SMC dynamics *in vivo*.

**Abstract**

The SMC complex, MukBEF, is important for chromosome organization and segregation in *Escherichia coli.* Fluorescently tagged MukBEF forms distinct spots (or 'foci') in the cell, where it is thought to carry out most of its chromosome associated activities. This chapter outlines the technique of Fluorescence Recovery After Photobleaching (FRAP) as a method to study the properties of YFP-tagged MukB in fluorescent foci. This method can provide important insight into the dynamics of MukB on DNA and be used to study its biochemical properties *in vivo*.

**1. Introduction**

The bacterial chromosome is compacted nearly a 1000-fold into a cell where it is faithfully replicated, transcribed and segregated. Not only is it highly compacted, it is also spatially organized with chromosomal regions occupying specific positions inside the cell *(1, 2)*. The highly conserved Structural Maintenance of Chromosome (SMC) complex, MukBEF, plays a central role in *E. coli* to maintain chromosome organization and ensure faithful chromosome segregation *(3, 4)*. The MukBEF complex consists of three proteins: The SMC-like MukB and two accessory proteins MukE and MukF. Deletion of any of these components results in a Muk$^-$ phenotype that includes temperature sensitivity, production of anucleate cells and loss of wild type chromosome organization. Functional fluorescent fusions of MukB, E or F all form foci in cells (Fig. 1), with two foci on average around the origin of replication *(3, 5)*. Recent studies using an array of microscopy-based approaches and genetic tools have provided insight into the properties of Muk foci and have supported the idea that foci are the centers of activity of the MukBEF complex *(3–7)*. The techniques used in these studies are widely applicable to understanding the functions of other proteins/ protein complexes *in vivo*.

In general, advances in live-cell imaging in combination with the use of Green Fluorescent Protein (GFP) and its variants have facilitated the ability to study the composition and dynamics of protein complexes in a cellular context *(8–19)*. In this chapter we describe the techniques of Fluorescence Recovery After Photobleaching (FRAP)*(20)* and Fluorescence Loss In Photobleaching (FLIP) to study the dynamics of YFP-tagged MukB, E or F in foci*(6)*. During FRAP, a subset of fluorescent molecules (typically, molecules in one of the two Muk foci) are irreversibly photobleached using a laser beam with high intensity of illumination. After the brief pulse of bleaching, images are recorded for subsequent time-frames at lower laser intensities to observe the recovery of fluorescence at the bleached spot either by diffusion of the non-bleached molecules or by active exchange of bleached molecules in the focus with unbleached ones. Since MukBEF forms two fluorescent foci, we can also record the loss in fluorescence of the unbleached focus (FLIP) during the photobleaching experiment. In an ideal scenario, the rate of recovery after photobleaching should be comparable to the loss in intensity of the unbleached focus.

This chapter will briefly describe the method to grow cells and prepare slides, similar to that described previously*(21)* and will outline a typical FRAP experiment as well as a simple method of data analysis. As states earlier, the method can be modified to study other protein complexes as well. For these experiments, *E. coli* cells are grown under conditions that result in non-overlapping replication cycles, so each cell has two Muk foci on average.

## 2. Materials

Instructions for the construction of YFP-tagged MukBEF components are beyond the scope of this chapter. However, a note is included on strain construction that might be useful (Note1).

### *2.1 Growth media:*

1. Luria Broth: 10g Yeast Extract, 5g NaCl and 5g Tryptone in 1L of water. pH is adjusted to 7 and LB is sterilized by autoclaving.

2. 10XM9 salt solution: 63g $Na_2HPO_4$, 30g $KH_2PO_4$, 5g NaCl and 10g $NH_4Cl$ in 1L of water. Sterilize by autoclaving.

3. 1XM9-glycerol: 1XM9 salts, 0.5 mg/ml of thiamine, 0.1% 1M $MgSO_4$, 0.1% 100mM $CaCl_2$ and 0.2% of glycerol as carbon source in water. Make 100mL of this solution.

4. 2XM9-glycerol: Same as 1XM9-glycerol but in half the quantity of water. Make 50mL of this solution.

5. 2% agarose: Invitrogen ultrapure agarose can be used for this. For 50mL, add 1 g of agarose to 50mL of water.

6. 1% agarose + M9-glycerol (this mixture is used for microscope slide preparation): Mix melted 2% agarose with 2XM9-glycerol in a 1:1 ratio. I usually mix 500μL of each solution by pipetting in an eppendorf and immediately use this to prepare the microscopy slide.

*2.2 Slides and microscope:*

1. Microscope slides: VistaVision microscope slides (VWR).

2. Coverslips: Micro cover glasses, Thickness 1.5, 24x50mm (VWR).

3. Gene Frame Seals. (Thermo Scientific Catalog number AB0578).

4. Microscope: UltraView PerkinElmer Spinning Disc Confocal microscope with FRAP module, 100x 1.35NA oil immersion objective, Electron-multiplying charge-coupled device (ImagEM, Hamamatsu Photonics) and UltraView PK Bleaching Device for photobleaching. Assuming that you will be imaging YFP-tagged MukBEF, the microscope should have laser lines for 514nm (see Note 2 for alternative laser lines).

5. Immersion oil: Immersol W oil NA 1.339 (Zeiss).

6. Software: Volocity imaging software (PerkinElmer) for image acquisition and ImageJ for image analysis.

**3. Methods:**

*3.1 Preparation of bacterial cultures for microscopy*

1. Streak bacterial cells from a frozen stock on LB agar plates with appropriate antibiotic at 37°C. As far as possible, use fresh cells no more than 2 weeks old. All cultures are grown by shaking to provide sufficient aeration. Most cells can grow at 37°C. (See Note 3 about growing *ΔmukBEF* cells). The steps listed below are for a strain carrying a YFP-tagged version of MukB. The same procedure can be followed for other tagged components of the complex.

2. Pick a single colony from the plate prepared in step1 and resuspend it in 5mL of LB. Allow the culture to grow until stationary phase (5-6 hr).

3. Make a 1 in 5000 dilution of the above into 5mL of 1XM9-glycerol and allow this culture to grow overnight.

4. The following day, subculture the cells in fresh 1XM9-glycerol (~1 in 1000 dilution) and allow the cells to grow till OD 0.1-0.2 (measured using a spectrophotometer). This should take 2-3 hr (see Note 4 for details about generation time of *E. coli* cells grown in M9-glycerol)

5. Spin down 500µL of culture from step 4 at 8000rpm for 1 min. Remove the supernatant and resuspend the pellet in 50µL of 1XM9-glycerol. Cells are now ready to be spotted on the microscope slide and imaged.

*3.2 Preparation of microscopy slide*

The procedure described here has been previously outlined in detail in another volume of this

series*(21)*. A condensed version of this protocol is provided below.

1. The Gene Frame is first stuck on a clean glass slide by removing its clear plastic cover. Make sure to stick the frame smoothly on all side without leaving wrinkles (see Note 5 on why Gene Frames are used).

2. Take 500µL of 1% agarose + M9glycerol and immediately transfer it to the centre of the gene frame prepared in the above step (See Note 6).

3. Place a coverslip on top of this and press it down to remove excess agarose and flatten the solution evenly in the frame. Let this stand for a few minutes, until the agarose has dried and solidified.

4. Once the agarose has solidified, slide the coverslip off and let the agarose dry for a couple of extra minutes.

5. Take 5µL of culture prepared earlier (step 5, section 3.1) and spot it on the agarose. Try to evenly distribute it across the slide by applying multiple spots and tilting the slide to allow spreading. Allow the slide to dry for a couple of more minutes. It is essential to do so as excess water will hamper step 6 of this section.

6. Remove the top plastic cover of the gene frame. On the sticky side of the frame carefully place a coverslip. Make sure that the coverslip is placed evenly and avoid the formation of air pockets. Once the coverslip has made contact with all four sides of the frame, you can press it down gently to even out its adhesion.

*3.3 Microscopy*

1. Turn on the lasers, microscope and computer. Then turn on Volocity (the acquisition software).

2. Add a drop of immersion oil to the objective and place the slide on top of the lens.

3. Cells should be focused using Brightfield or DIC. Avoid focusing using fluorescence to prevent photobleaching. An ideal field of view for imaging should have cells evenly distributed and in focus. A typical field can have up to 50 cells.

4. Open the settings for YFP (514 nm laser) on Volocity and reduce laser power to 4-6% (See Note 7). Under camera settings, set the frame rate to 300 ms for image capture. Focus and take a picture.

5. In the picture, you will be able to see typically two distinct MukB-YFP foci per cell. The aim of the experiment is to bleach one of the two foci and record fluorescence recovery after bleaching (See Note 8 on use of cephalexin to elongate cells).

6. Open the FRAP module to set up bleaching conditions (See Note 9 on FRAP calibration). Pulse bleach is ideally done with 6-15% laser intensity for 15 ms. The number of cycles of bleaching is limited to 1. A region of interest (ROI) is drawn around the focus to be bleached. This is usually a diffraction limited region of ~300 nm (See Note 10 on size of ROI).

7. Using the PhotoKinesis menu, choose up to six ROIs (one ROI per cell) in one field of view. ROIs can be chosen by drawing a region around a MukB-YFP focus. Then set the conditions for acquisition. Typically, take 2-3 pre-bleach images and after pulse-bleaching (step 6), record recovery of fluorescence every 15 sec for 3 min or every 30 sec for 5 min. Again, image capture should be done at lower laser intensity (4-6%) at a 300 ms capture rate. The entire module is automated. Once the settings have been applied and acquisition has started, images will be acquired in the sequence desired: two pre-bleached images, followed by pulse-photobleaching of the ROIs selected, followed by image capture with low laser intensities for 3 or 5 min. Movies are saved as stack files that can be opened in ImageJ.

8. Repeat the above procedure after moving to a new field of view that is distinct from the field previously imaged (See Note 11).

*3.4 Image analysis*

1. Open images (saved as a stack) in ImageJ. It is important to remove background fluorescence prior to extracting information on focus intensity. This is done using the background subtraction module in ImageJ. Apply subtraction to the entire stack.

2. For FRAP measurements draw a region of interest around the spot that was bleached in the experiments in section 3.3. Also draw a second ROI around the entire cell to calculate total cellular fluorescence intensity for the cell undergoing pulse-bleaching. Use ImageJ's 'Measure Intensity' tool to extract mean and total intensity values for each ROI through the entire stack.

3. For FLIP measurements, the same procedure (step 2) should be repeated for an ROI drawn around the unbleached focus in the cell undergoing pulse-bleaching.

4. FRAP, FLIP and total cellular fluorescence intensities for a cell in a movie can now be copied and pasted into Excel. To compare recovery across cells, intensity of ROIs should be normalized to highest pre-bleach intensities.

5. Before calculating recovery times, it is important to correct for photobleaching due to fluorescence excitation during imaging. This is done by normalizing total cellular intensity at each time point to the total cellular intensity soon after photobleaching.

6. Now the intensity of a bleached focus at a given time point can be calculated using the following equation, which corrects measured intensity values for any photobleaching which may have occured:

$I(t) = (Ib(t)/Ib_{max}) / (Ic(t)/Ic_{max})$

Where:

$Ib(t)$ = intensity of ROI at time $t$ (post bleach).

$Ib_{max}$ = maximum intensity of ROI (pre bleach).

$Ic(t)$ = intensity of whole cell at time $t$.

$Ic_{max}$ = intensity of whole cell soon after bleach.

7. By plotting the $I(t)$ values for a bleached or unbleached focus over the time of imaging, you can get an estimate of FRAP or FLIP respectively (See Note 12 on expected outcomes and controls) (Fig. 2).

## 4. Notes

1. It is ideal to construct fluorescent fusions of proteins in the chromosome at the endogenous locus of the gene. One efficient way of strain construction in *E. coli* is using the λ-Red recombination system*(22)*. MukBEF genes are arranged in an operon (in the order *mukF-mukE-mukB*). While C-terminal fusions to MukB and MukE are fully functional, MukF needs to be tagged in its N-terminus for function to be maintained. A short linker of about 8-10 amino acids (Glycine, serine and alanine rich) is typically inserted between MukB, E or F and the fluorescent protein (monomeric form of YFP, mYPet, has been used to image MukBEF in previous experiments*(6)*).

2. Typically pulse-bleaching should be carried out using the same wavelength as used for imaging. In the case of YFP this is the peak absorption wavelength of 514nm. In the event that the YFP laser is not powerful enough for pulse-bleaching, a 488nm laser can be used for this step of the experiment.

3. Wild type *E. coli* cells can be grown at 37°C in rich or minimal media. However, *ΔmukB, E or F* strains or strains with mutants of MukB are temperature sensitive and ideally grow at room temperature (~22°C). When doing an experiment that involves *ΔmukB* (or mutant MukB) and wild type cells, you should grow both cultures at 22°C so that the conditions are comparable during imaging as well.

4. The generation time for *E. coli* in M9-glycerol is ~100min. Cells grown in these conditions have non-overlapping replication cycles and are simpler to study processes such as chromosome organization, replication and segregation using microscopy. When growing cells in M9-glycerol, it is important to ensure that cells do not go into late stationary phase ($O.D_{600}$~1) as the recovery time (lag phase) to return to exponential growth will be prolonged.

5. Agarose pads can dry out or dessicate when kept for a long period during imaging (especially at high temperatures such as 37°C). In order to prevent this, gene frames are used.

6. As stated earlier, M9-glycerol provides ideal growth conditions for microscopy-based experiments in *E. coli.* Another important advantage of using M9-glycerol over LB is the lower auto fluorescence in M9. Background fluorescence can pose a problem during imaging and in particular, during analysis of fluorescence intensity in the cell. It is always advisable to use media with low levels of auto fluorescence for this reason. Auto fluorescence can be further reduced using low fluorescence agarose, for example the Nusieve GTG Agarose from Lonza Biosciences.

7. Ideal laser intensity settings will vary between microscope setups. It is recommended to use low laser intensities during imaging in order to reduce photobleaching or phototoxicity effects. The same applies for laser intensity used for pulse-bleaching. It is recommended to use an intensity that is high enough to completely bleach fluorescence in the region of interest, but not too high that other regions of the cell are bleached as well.

8. Since *E. coli* cells are small in size, FRAP experiments can sometimes cause bleaching across the entire cell, which can complicate analysis of fluorescence recovery. One way of circumventing this problem is to treat *E. coli* cells with the cell division inhibitor cephalexin (100mg/mL) for 2-3 generations prior to imaging. This will result in the production of elongated cells with multiple, segregated chromosomes. If cephalexin is used, it should also be added to the agarose pad to prevent cells from dividing during imaging.

9. Before you start a FRAP experiment it is important to ascertain that the pulse-bleach is centered on the region of interest chosen in the cell. This can be done using the 'FRAP Calibration Wizard' in Volocity. For this, you will need a slide with GFP fluorescence (I use a fluorescence marker to make this).

10. Since bacterial cells are small, try to use a small ROI for pulse-bleaching to avoid bleaching a large area of the cell.

11. While imaging, it is important to ensure that cells are still actively growing on the agarose pad. I recommend FRAP imaging of cells on an agarose pad for no longer than 2 hrs. More traditional time-lapse movies can be carried out for longer (as long as the cells continue to grow).

12. There are, broadly, two typical outcomes of a FRAP experiment: a. there is no recovery after photobleaching and the slight increase in fluorescence intensity in the ROI after pulse-bleaching is due to diffusion of free fluorescent molecules into the area. b. there is active recovery of fluorescence as assessed by a significant increase in intensity in the ROI after pulse-bleaching. You should be able to see the return of a MukB focus in this case. In order to test for the physiological relevance of this recovery, you can use MukB mutants that should not show recovery after photobleaching*(6)*.

**Acknowledgements**

A.B. is supported by a Human Frontiers Science Program Postdoctoral Fellowship.

**Figure legends:**

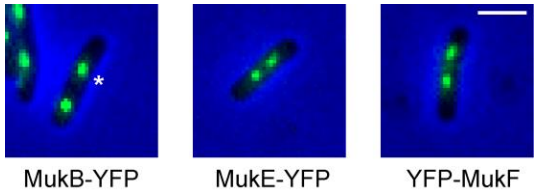

**Fig. 1: Fluorescent fusions of MukBEF form foci in cells.**

YFP-tagged MukB, MukE and MukF form foci in cells. Representative cells are shown in this figure. Fluorescent focus is highlighted with *.

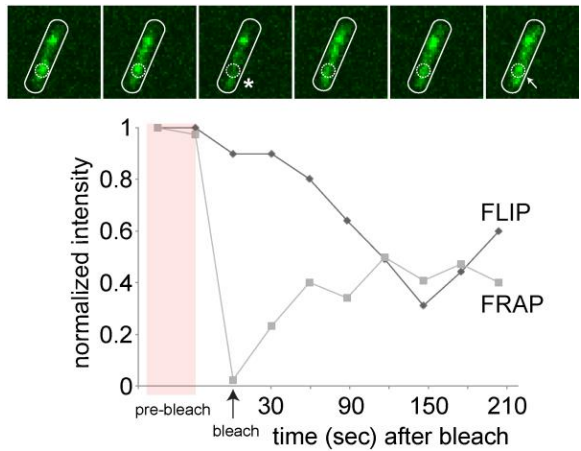

**Fig. 2: Using FRAP to study dynamics of MukB-YFP in foci.**

Above: Representative time-lapse of a cell with MukB-YFP foci during a FRAP experiment is shown. The region of interest (ROI) that is pulse-bleached is highlighted with a circle, pulse-bleaching is indicated with * and recovery after bleaching is indicated by the arrow. Below: Quantification of FRAP experiment is shown. Two pre-bleach images were taken prior to pulse-bleaching of fluorescence in the ROI. Images were taken every 30 sec after bleaching to record fluorescence recovery after bleaching. Normalized intensity in plotted for the bleached focus (FRAP) and for the control, unbleached focus (FLIP) in the same cell.